\documentclass[11pt]{article}
\usepackage{epsfig}
\usepackage{cite}
\input{epsf}
\newcommand{\beq}{\begin{equation}}
\newcommand{\eeq}{\end{equation}}
\newcommand{\beqa}{\begin{eqnarray}}
\newcommand{\eeqa}{\end{eqnarray}}
\newcommand{\eq}{\begin{eqnarray}}
\newcommand{\en}{\end{eqnarray}}

\def\md0{\stackrel{0}{m}_\Delta}
\def\mN0{\stackrel{0}{m}_N}

\begin{document}

\begin{flushright}
{\tiny{FZJ-IKP-TH-2007-01}}
{\tiny{HISKP-TH-07/01}}
\end{flushright}

\vspace{.6in}

\begin{center}

\bigskip

{{\Large\bf The $\Delta$-resonance in a finite volume}\footnote{This
research is part of the EU Integrated Infrastructure Initiative Hadron 
Physics Project under contract number RII3-CT-2004-506078. Work supported 
in part by DFG (SFB/TR 16,``Subnuclear Structure of Matter'') and by the 
EU Contract No. MRTN-CT-2006-035482, \lq\lq FLAVIAnet''.}}

\end{center}

\vspace{.3in}

\begin{center}
{\large 
V\'eronique Bernard$^a$\footnote{email: bernard@lpt6.u-strasbg.fr},
Ulf-G. Mei{\ss}ner$^{b,c}$\footnote{email: meissner@itkp.uni-bonn.de}
{\footnotesize and}
Akaki Rusetsky$^{b,d}$\footnote{email: rusetsky@itkp.uni-bonn.de}
}

\vspace{1cm}

$^a${\it Universit\'e Louis Pasteur, Laboratoire de Physique
            Th\'eorique\\ 3-5, rue de l'Universit\'e,
            F--67084 Strasbourg, France}

\bigskip

$^b${\it Universit\"at Bonn,
Helmholtz--Institut f\"ur Strahlen-- und Kernphysik (Theorie)\\
Nu{\ss}allee 14-16,
D-53115 Bonn, Germany}

\bigskip

$^c${\it Forschungszentrum J\"ulich, Institut f\"ur Kernphysik 
(Theorie)\\ D-52425 J\"ulich, Germany}

\bigskip

$^d\,${\it On leave of absence from: High Energy Physics Institute,\\
Tbilisi State University,
University St.~9, 380086 Tbilisi, Georgia}

\bigskip

\bigskip

\end{center}

\vspace{.4in}

\thispagestyle{empty} 

\begin{abstract}\noindent 
We study the extraction of  $\Delta$-resonance parameters from
lattice data for small quark masses, corresponding to the case of an
unstable $\Delta$. To this end, we calculate the spectrum of the correlator of 
two $\Delta$-fields in a finite Euclidian box up-to-and-including 
$O(\epsilon^3)$ in the small scale expansion using
 infrared regularization. On the basis of 
our numerical study, we argue that the extraction of the parameters of 
the $\Delta$-resonance (in particular, of the mass and the 
pion-nucleon-delta 
coupling constant) from the measured volume dependence of the lowest energy 
levels should be feasible. 
\end{abstract}

\vspace{1cm}
\footnotesize{\begin{tabular}{ll}
{\bf{Pacs:}}$\!\!\!\!$&  12.38.Gc, 12.39.Fe, 11.10.St
\\
{\bf{Keywords:}}$\!\!\!\!$& Lattice QCD, Baryon resonances, Chiral Lagrangians,
Field theory in a finite volume\\
\\
\end{tabular}} 

\vfill

\pagebreak

\section{Introduction}
\label{sec:intro}

\normalsize

Monte-Carlo simulations in lattice QCD enable one
 to calculate the spectrum
of low-lying hadrons from first principles and thus to obtain important 
information about the long-range
dynamics of quarks and gluons. In the past, a major effort has been 
concentrated on the study of the ground-state
 particle spectrum. However, recently
there has been a lot of work on the spectroscopy of the excited nucleon 
states as
well~\cite{Maynard:2002ys,Richards:2001bx,Gattringer:2003qx,Sasaki:2003xc,Sasaki:2001nf,Zanotti:2001yb,Melnitchouk:2002eg,Zhou:2006xe,Sasaki:2005ug}.
For the status of lattice calculations of the baryon spectrum, 
see e.g. the recent reviews~\cite{McNeile:2003dy,Leinweber:2004it}. 
We note that several long-standing puzzles in the field are still
awaiting a resolution. In particular, it is worth to mention
the problem of the level ordering of the negative-parity nucleon excitation
$N^*\,(1535)$ and the positive-parity Roper resonance $N^*\,(1440)$ 
as well as the structure of the $\Lambda\,(1405)$.

The $\Delta$-resonance is the most important baryon resonance. Its mass
is close to the mass of the nucleon, and it couples strongly to nucleons, 
pions and photons. It is clear that a systematic study of the properties of the
$\Delta$-resonance in lattice QCD could lay a solid theoretical basis for 
understanding the low-energy QCD dynamics in the one-baryon sector in lattice
QCD.
Evaluating the mass of the $\Delta$-resonance has already been addressed 
in the last few years. For illustration, we consider the recent
calculation of the $\Delta$-mass in quenched QCD using a tadpole-improved
anisotropic action and FLIC fermions~\cite{Zhou:2006xe}.
In this paper, the local interpolating field 
with quantum numbers of the $\Delta^+$
has been chosen in the following manner
\eq\label{eq:chi}
\chi_\mu=\frac{1}{\sqrt{3}}\,
\epsilon_{abc}\,\biggl\{
2\bigl(u^{aT}C\gamma_\mu d^b\bigr)u^c
+\bigl(u^{aT}C\gamma_\mu u^b\bigr)d^c\biggr\}\, ,
\en
where $a,b,c$ denote color indices and $C$ is the charge conjugation matrix.
With this interpolating field one further calculates the two-point correlation
function at zero spatial momentum and at large Euclidean time,
extracting the mass spectrum. More precisely, 
in Ref.~\cite{Zhou:2006xe} the calculation of the ratio
 $m_\Delta/m_N$ (as well as the similar ratios for some other excited
baryon states) has been performed
 down to the quark masses corresponding to
$(M_\pi/M_\rho)^2=0.4$. At such quark masses, the dependence of the
data on $(M_\pi/M_\rho)^2$ is very smooth (almost linear). However, 
a large gap in the quark (pion) mass has still to be bridged if the data are
extrapolated down to the physical value of $(M_\pi/M_\rho)^2 \simeq 0.03$.

The above example highlights all features of the lattice calculations
of the parameters of baryon resonances (and the
$\Delta$-resonance, in particular), which we wish to put
under scrutiny in the present paper:
\begin{itemize}
\item[a)]
As already mentioned, the distance from the lowest data point to the
physical value of the quark mass is still very large and this will lead
to large extrapolation errors. In case of a stable
particle, one would argue that performing calculations at smaller
quark masses will reduce this uncertainty until, eventually, the
simulations are done close to or at the physical value of the quark mass. However,
in the case of a resonance, the threshold value of the quark mass
exists after which e.g. the $\Delta$ starts to decay into a pion
and a nucleon. It is evident that, 
below the threshold, the method which was described above
can not be applied straightforwardly.
Does this mean that, in the case of a resonance, there is
{\em always} a (not so small) gap in the lattice data, where one
should rely only on chiral extrapolation? 
\item[b)]
Lattice data are always real. Does this mean that one gets 
the real part of the resonance pole mass as a result of a chiral
extrapolation below the decay threshold? 
\item[c)]
Can one determine the decay width of a resonance by combining the method
described above with the chiral extrapolation?
\end{itemize} 

Note also that the Monte-Carlo simulations
e.g. in Ref.~\cite{Zhou:2006xe} have been carried out in the quenched
approximation. It is however clear that  unstable systems at small
quark masses, which will be considered below, can be meaningfully discussed
only in the context of  lattice data based on simulations with dynamical fermions.

To summarize, we want to ask whether there
exists a modification of the above method that enables one 
(at least, in principle) to carry out 
calculations of the properties of an excited state at such quark masses
when this state becomes unstable. Moreover, one would like
 to eventually walk all way
down to the physical quark mass, excluding the extrapolation error
altogether (like this appears possible for ground-state hadrons). It would also
be very instructive to see in detail what happens in the
vicinity of the threshold, when one crosses it
while performing a chiral extrapolation of the spectrum.
Finally, one should investigate which quantities are most sensitive to the
resonance parameters at small quark masses, and how accurately one could
extract the mass and the width of the resonance by calculating these
quantities on the lattice.

\begin{figure}[t]
\begin{center}
\includegraphics[width=9.cm]{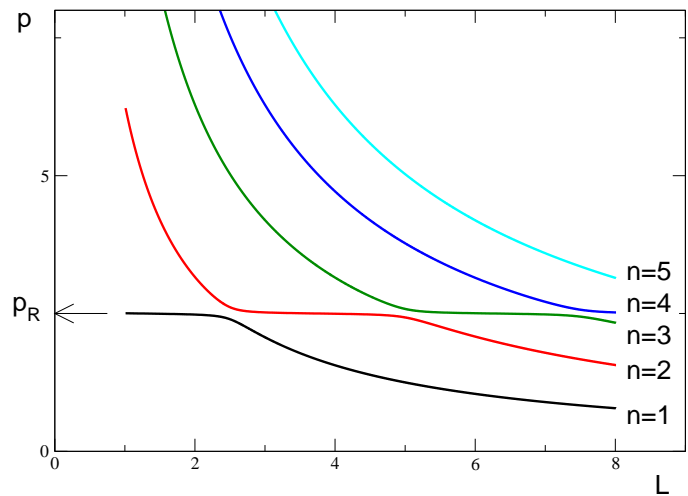}
\end{center}
\caption{A schematic representation of the avoided level crossing 
in the presence of a narrow resonance. The center-of-mass momentum of a two-particle
pair is plotted against the size of a box $L$ (arbitrary units). It is seen
that in the vicinity of the resonance momentum $p_R$ a peculiar behavior of
the energy levels is observed. The exact position of the resonance in this case
could be easily pinned down by measuring the energy levels on the lattice.}
\label{fig:avoided}
\end{figure}

The question of identification of hadron resonances on the lattice has
already been addressed in the past. For example, we would like to mention
the papers~\cite{LuescherII,Luescher_torus,Luescher_rho,Houches,Wiese,Rummukainen:1995vs,Christ:2005gi,Kim:2005gf}. In general, the method
 proposed originally by L\"uscher considers the
extraction of the two-body scattering phase shifts from the energy levels,
calculated in a finite Euclidean box. In particular, 
the signatures of  unstable particles have been studied. It is 
demonstrated that in the presence of a narrow resonance the dependence
of the energy spectrum of the system on the box size $L$ exhibits a very 
peculiar behavior near the resonance energy, where the so-called ``avoided 
level crossing'' takes place, see Fig.~\ref{fig:avoided}.
 Note that usually this name
is used to describe an abrupt rearrangement of the structure of
the energy levels of a system, which takes place near the resonance energy 
when the scattering phase passes through $\pi/2$. From  Fig.~\ref{fig:avoided}
it is clear that the position of a narrow resonance can be readily 
identified by measuring the energy levels of a system at a finite volume and
locating horizontal ``plateaus.''
Moreover, the minimal distance that separates the energy levels near the
avoided crossing, is determined by the decay width of a resonance and,
consequently, the latter quantity can be also extracted from the same lattice
data. A very nice qualitative discussion of the avoided level crossing
is given e.g. in Refs.~\cite{Houches,Wiese,DeGrand}.

An approach, related to the L\"uscher's method,
is used in Ref.~\cite{DeGrand} to study
 the structure of the energy levels in the two-pion system carrying the
quantum numbers of the $\rho$-meson. To this end, 
these energy levels are calculated in
an effective field theory (EFT) with a phenomenological lowest-order 
$\rho\pi\pi$ coupling.
Furthermore, in Ref.~\cite{Michael} it has been shown that the presence of a 
narrow excited state above the threshold modifies the simple exponential decay
law of the time-sliced two-point function. The decay width
within this approach is extracted not from the two-point function, but 
directly from the decay amplitudes (see also~\cite{Loft:1988sy,Lellouch:2000pv}). In addition, we point out the recent 
investigation~\cite{Yamazaki}, where it is proposed to reconstruct the 
spectral density in the two-point function by using the maximum entropy method.
This approach, in principle, also has the capability to address the
problem of unstable states in lattice calculations.

Note also that  L\"uscher's approach has been recently applied 
to study nucleon-nucleon phase shifts at low-energy, as well as two-body 
shallow bound 
states~\cite{Beane:2003yx,Beane:2003da,Beane:2006gf,Beane:2006mx,Sasaki:2006jn}.
To this end, in Ref.~\cite{Beane:2003yx}, L\"uscher's master formula 
which relates the scattering
phase shifts to the energy level displacements caused by the $NN$ interaction,
is re-derived within a non-relativistic EFT. Note that
at large box sizes, which are used in the study of the scattering processes,
the characteristic center-of-mass momenta are small and hence working within
 the non-relativistic EFT is justified.

An approach, which we use in the present paper, is closely related 
to the one of 
Refs.~\cite{LuescherII,Luescher_torus,Luescher_rho,Houches,Wiese,DeGrand}.
To be precise, we investigate the energy levels of the system with the quantum numbers
of the $\Delta$-resonance in a finite box with the size $L$. This is achieved by
 calculating the self-energy of the $\Delta$ in the small scale expansion 
(SSE)\footnote{The SSE is a phenomenological extension of chiral
  perturbation theory in which the nucleon-$\Delta$ mass splitting is counted
 as an additional small parameter. This quantity, however, does not vanish in
 the chiral limit. The framework of the SSE is laid out in detail in
 Ref.~\cite{Hemmert:1997ye}}
at a finite volume and finding the poles of the propagator, which 
yields the parameterization of the spectrum of the Hamiltonian in a 
finite Euclidean box
as a function of the variables $M_\pi$ and $L$, as well as the physical masses
$m_\Delta(M_\pi)$, $m_N(M_\pi)$ and the $\pi N\Delta$ coupling constant,
denoted $g_{\pi N \Delta}$. We further investigate 
the behavior of energy levels with respect to these parameters. 
The main question which is addressed here is
whether the dependence of the energy levels on the above-mentioned parameters
is pronounced enough in order to enable one to perform an accurate fit to  
determine the mass and width of the $\Delta$-resonance.
Here we would like to mention that, as it turns out later, the width of 
$\Delta$ is so large
that the avoided level crossing is almost completely washed out from the 
energy spectrum. Therefore, the nice procedure shown in Fig.~\ref{fig:avoided}
does not immediately apply and,
in order to be able to accurately extract the parameters of the 
$\Delta$-resonance even in this case, one has to find an optimum 
fitting strategy. The discussion of this issue
constitutes  the main content of this work.

The paper is organized as follows. In section~\ref{sec:formalism} we consider
the formalism that is used to calculate the self-energy of the 
$\Delta$-resonance in the SSE, including a discussion 
of the infrared regularization procedure in a  finite volume.
Section~\ref{sec:numerics} is devoted to the numerical calculations and the
description of the fitting algorithm and the presentation of our
results. In section~\ref{sec:analytic} we introduce 
analytic parameterizations
of the energy spectrum at finite volume, which might be useful for 
performing 
fits to the lattice data. Finally, section~\ref{sec:concl} contains our conclusions.

\section{The formalism}
\label{sec:formalism}

\subsection{Extraction of the energy spectrum}

Let us consider the two-point function of two
$\Delta$--fields, given by Eq.~(\ref{eq:chi}), in a Euclidean
 box and choose the rest frame.
After projecting out the spin-$\frac{3}{2}$ state,
this two-point function in  momentum space develops poles at 
$p_\mu^n=(iE_n,{\bf 0})$. In  coordinate space, the same two-point function
at large Euclidean times $t$ will be given by a sum of exponentials
\eq
\sum_{{\bf x}}\langle 0|\chi_\mu(x)\bar\chi_\nu(0)|0\rangle_L
\to \sum_n (z_n)_{\mu\nu}\exp(-E_nt)\, .
\en 
where the subscript $L$ indicates that the calculations are done 
at a finite volume $L\times L\times L \times L_t$.  We further 
assume that the size of the box in the time direction $L_t$ is
much larger than in the spatial direction $L$ 
(eventually, $L_t\to \infty$)
Consequently, studying the behavior of this two-point function
at large Euclidean times, one is able to extract the energy levels
$E_n=E_n(L)$ of the positive parity, spin-$\frac{3}{2}$ states,
which are functions of the box size $L$. 
In case of a stable $\Delta$-state, finite-size corrections to the
lowest energy level vanish exponentially, 
$E_1(L)-m_\Delta=\exp(-\mbox{const}\cdot L)$ and in the large-$L$ limit
this level yields the value of the stable $\Delta$-mass. This is, however, not
the case for the decaying $\Delta$, when the dependence of the energy levels
on $L$ is governed by a power rather than by an exponential law.
The question is, whether one can extract the parameters of the 
$\Delta$-resonance from the measured dependence of $E_n(L)$ on $L$.

\subsection{The Lagrangian and Feynman rules}

In order to parameterize the volume-dependent
energy levels of the system in terms of the
$\Delta$-resonance parameters, 
we calculate  the same two-point function in the SSE at finite volume. 
These calculations are similar to the recent study of
the volume-dependent nucleon mass~\cite{Detmold:2004ap,Bedaque:2004dt,AliKhan:2003cu,Beane:2004tw,Colangelo:2005cg}
-- except for the fact that in the present paper we deal with an
unstable particle. The calculations are performed
by using the effective chiral Lagrangian, which
explicitly contains pion, nucleon and $\Delta$ degrees of 
freedom~\cite{Hemmert:1997ye,Bernard:2005fy} and coincides with the Lagrangian
used at infinite volume. The relevant terms (in Minkowski space) are
displayed below:
\eq\label{eq:L}
{\cal L}=\frac{F^2}{4}\,\langle \partial^\mu U\partial_\mu U^\dagger
+\chi^\dagger U+U^\dagger\chi\rangle+\cdots+{\cal L}_{\pi N}+
{\cal L}_{\pi \Delta}+{\cal L}_{\pi N\Delta}\, ;
\en
\eq\label{eq:LN}
{\cal L}_{\pi N}&=&{\cal L}_{\pi N}^{(1)}+{\cal L}_{\pi N}^{(2)}
+{\cal L}_{\pi N}^{(3)}+\cdots
=\bar\psi_N \biggl\{
\Lambda_{\pi N}^{(1)}+\Lambda_{\pi N}^{(2)}+\Lambda_{\pi N}^{(3)}
+\cdots\biggr\}
\psi_N\, ,
\nonumber\\[2mm]
\Lambda_{\pi N}^{(1)}&=&i\not\!\! D\,-\mN0+\frac{g_A}{2}\,\not\! u\,\gamma_5\, ,
\nonumber\\[2mm]
\Lambda_{\pi N}^{(2)}&=&c_1\langle\chi_+\rangle-\frac{c_2}{4(\mN0)^2}\,
\biggl(\langle u_\mu u_\nu \rangle D^\mu D^\nu+\mbox{h.c.}\biggr)
+\frac{c_3}{2}\, \langle u_\mu u^\mu\rangle+\cdots\, ,
\nonumber\\[2mm]
\Lambda_{\pi N}^{(3)}&=&
B_1^N\Delta_0\langle\chi_+\rangle+B_0^N\Delta_0^3+\cdots\, ;
\en

\eq\label{eq:LND}
{\cal L}_{\pi N\Delta}&=&{\cal L}_{\pi N\Delta}^{(1)}+\cdots\, ,\quad\quad
{\cal L}_{\pi N\Delta}^{(1)}=g_{\pi N\Delta}\bar\psi^i_\alpha O^{\alpha\beta}
w^i_\beta\psi_N+\mbox{h.c.}\, ;
\en

\eq\label{eq:LD}
{\cal L}_{\pi\Delta}&=&{\cal L}_{\pi\Delta}^{(1)}+{\cal L}_{\pi\Delta}^{(2)}
+{\cal L}_{\pi\Delta}^{(3)}+\cdots
=-\bar\psi^i_\alpha O^{\alpha\mu}\biggl\{
(\,\Lambda_{\pi\Delta}^{(1)}\,)^{\mu\nu}_{ij}
+(\,\Lambda_{\pi\Delta}^{(2)}\, )^{\mu\nu}_{ij}
+(\,\Lambda_{\pi\Delta}^{(3)}\, )^{\mu\nu}_{ij}
+\cdots\biggr\}O^{\nu\beta}\psi_\beta^j\, ,
\nonumber\\[2mm]
(\,\Lambda_{\pi\Delta}^{(1)}\,)^{\mu\nu}_{ij}&=&
O^{\mu\alpha}\biggl(g_{\alpha\beta}(i\! \not\!\! D_{ij}\,
-\md0\xi^{3/2}_{ij})-\frac{1}{2}\,
\{\gamma_\alpha\gamma_\beta,(\! i\not\!\! D_{ij}\,
-\md0\xi^{3/2}_{ij})\}\biggr)O^{\beta\nu}
+\frac{g_1}{2}\,\not\! u^{ij}\,\gamma_5 g^{\mu\nu}\, ,
\nonumber\\[2mm]
(\,\Lambda_{\pi\Delta}^{(2)}\,)^{\mu\nu}_{ij}&=&
g^{\mu\nu}\biggl\{a_1\langle\chi_+\rangle\delta_{ij}-\frac{a_2}{4(\md0)^2}\,
\biggl(\langle u_\alpha u_\beta\rangle D^\alpha_{ik} D^\beta_{kj}+\mbox{h.c.}
\biggr)
+\frac{a_3}{2}\,\langle u_\mu u^\mu\rangle\delta_{ij}\biggr\}+\cdots\, ,
\nonumber\\[2mm]
(\,\Lambda_{\pi\Delta}^{(3)}\,)^{\mu\nu}_{ij}&=&
g^{\mu\nu}\delta_{ij}\biggl\{B_1^\Delta\Delta_0\langle\chi_+\rangle
+B_0^\Delta\Delta_0^3\biggr\}+\cdots\, ,
\en
where $\langle\cdots\rangle$ denotes the trace in  flavor space.
Throughout, we work in the isospin limit $m_u = m_d = \hat m$. 
The building blocks for the above Lagrangian are defined by
\eq
&&U=u^2\, ,\quad\quad
u_\mu=iu^\dagger\partial_\mu U u^\dagger\, ,\quad\quad
D_\mu=\partial_\mu+\frac{1}{2}\, [u^\dagger,\partial_\mu u]\, ,
\nonumber\\[2mm]
&&\chi=2B(s+ip)\, ,\quad \chi_+=u^\dagger\chi u^\dagger+u\chi^\dagger u\, ,
\quad\quad s=\hat m {\bf 1}+\cdots\, ,
\nonumber\\[2mm]
&&w^i_\mu=\frac{1}{2}\,\langle\tau^i u_\mu\rangle\, ,\quad\quad
D^\mu_{ij}=\delta_{ij}D^\mu-i\epsilon_{ijk}\langle \tau^kD^\mu\rangle\, ,
\quad\quad
u^\mu_{ij}=\delta_{ij} u^\mu
\en
and
\eq
O^{\mu\nu}=g^{\mu\nu}-\frac{2}{d}\,\gamma^\mu\gamma^\nu\, .
\en
In these formulae, 
$U=\exp\bigl\{i\,\mbox{\boldmath ${\tau}$}\cdot\mbox{\boldmath${\pi}$}/F\bigr\}$, 
$\psi_N$ and $\psi_\mu^i$ represent the pion, nucleon and $\Delta$
 interpolating fields, respectively. Further, $\mN0$, $\md0$ are the masses
of the nucleon and the $\Delta$ in the chiral limit, and $\Delta_0=\md0-\mN0$
is the $\Delta$-nucleon mass difference in this limit. The pion mass
is given by $M_\pi^2=2B\hat m+O(\hat m^2)$. Furthermore, $F$, $g_A$, 
$g_{\pi N\Delta}$ denote the pion decay constant, the nucleon axial-vector 
constant and the $\pi N\Delta$ coupling constant in the chiral limit and
$c_i,B_i^N,a_i,B_i^\Delta$ denote various low-energy constants (LECs).
Finally, the projectors onto the isospin-$\frac{3}{2}$ and 
isospin-$\frac{1}{2}$ subspaces are defined by
\eq
\xi^{3/2}_{ij}=\delta_{ij}-\frac{1}{3}\tau_i\tau_j\, ,\quad\quad
\xi^{1/2}_{ij}=\frac{1}{3}\,\tau_i\tau_j\, ,\quad\quad
\xi^{3/2}_{ij}+\xi^{1/2}_{ij}=\delta_{ij}\, .
\en
The free $\Delta$-propagator in  $d$-dimensional space can be read off 
from the quadratic part of the Lagrangian
\eq\label{eq:Dprop0}
i\langle 0|T\psi_\mu^{0,i}(x)\bar\psi_\nu^{0,j}(0)|0\rangle
=\int\frac{d^dp}{(2\pi)^d}\,\mbox{e}^{-ipx}\,S^0_{\mu\nu}(p)
\,\xi^{3/2,ij}\, 
\en
with
\eq
S_{\mu\nu}^0(p)&=&-\,\frac{1}{\md0-\not\! p}\,
\biggl\{ g_{\mu\nu}-\frac{1}{d-1}\,\gamma_\mu\gamma_\nu
-\frac{(d-2)p_\mu p_\nu}{(d-1)(\md0)^2}
+\frac{p_\mu\gamma_\nu-p_\nu\gamma_\mu}{(d-1)\md0}\biggr\}
=-\,\frac{1}{\md0-\not\! p}\,(P^{3/2})_{\mu\nu}
\nonumber\\[2mm]
&+&\frac{1}{\sqrt{d-1}\,\md0}\,
\biggl\{(P_{12}^{1/2})_{\mu\nu}+(P_{21}^{1/2})_{\mu\nu}\biggr\}\
-\frac{d-2}{(d-1)(\md0)^2}(\not\! p\,+\md0)(P_{22}^{1/2})_{\mu\nu}\, ,
\en
where the projectors onto the spin-$\frac{3}{2}$ and spin-$\frac{1}{2}$ states
are defined as
\eq
(P^{3/2})_{\mu\nu}&\! =&g_{\mu\nu}-\frac{1}{d-1}\,\gamma_\mu\gamma_\nu
-\frac{1}{(d-1)p^2}\,(\not\! p\,\gamma_\mu p_\nu+p_\mu\gamma_\nu\!\not\! p\,)
-\frac{d-4}{d-1}\,\frac{p_\mu p_\nu}{p^2}\, ,
\nonumber\\[2mm]
(P_{12}^{1/2})_{\mu\nu}&\! =&\frac{1}{\sqrt{d-1}\,p^2}\,
(p_\mu p_\nu-\not\! p\,p_\nu\gamma_\mu)\, ,
\nonumber\\[2mm]
(P_{21}^{1/2})_{\mu\nu}&\! =&\frac{1}{\sqrt{d-1}\,p^2}\,
(\not\! p\,p_\mu\gamma_\nu-p_\mu p_\nu)\, ,
\nonumber\\[2mm]
(P_{22}^{1/2})_{\mu\nu}&\! =&\frac{p_\mu p_\nu}{p^2}\, ,
\nonumber\\[2mm]
(P_{11}^{1/2})_{\mu\nu}&\! =&g_{\mu\nu}-(P^{3/2})_{\mu\nu}
-(P_{22}^{1/2})_{\mu\nu}\, .
\en
These projectors obey the following relations
\eq
&&(P^{1/2}_{ij})_{\mu\lambda}(P^{1/2}_{kl})^{\lambda\nu}=\delta_{jk}
(P^{1/2}_{il})_\mu^\nu\, ,
\nonumber\\[2mm]
&&(P^{3/2})_{\mu\lambda}(P^{1/2}_{ij})^{\lambda\nu}
=(P^{1/2}_{ij})_{\mu\lambda}(P^{3/2})^{\lambda\nu}=0\, ,
\nonumber\\[2mm]
&&(P^{3/2})_{\mu\lambda}(P^{3/2})^{\lambda\nu}=(P^{3/2})_\mu^\nu\, ,
\quad\quad
i,j,k,l=1,2\, .
\en

\begin{figure}[t]
\begin{center}
\includegraphics[width=11.cm]{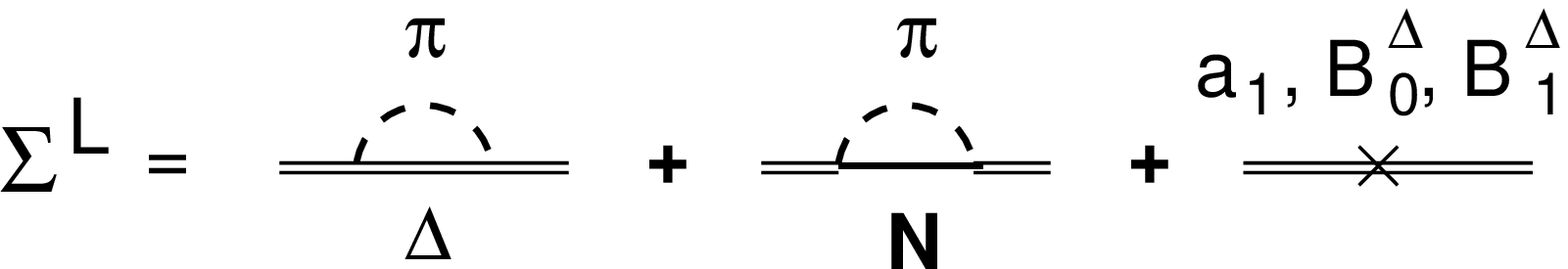}
\end{center}
\caption{The self-energy of the $\Delta$ at $O(\epsilon^3)$ in the SSE.}
\label{figure1}
\end{figure}

\subsection{The poles of the propagator}

The $\Delta$ propagator after inclusion of the self-energy diagrams
(see Fig.~\ref{figure1}) becomes
\eq\label{eq:DpropM}
i\langle 0|T\psi_\mu^i(x)\bar\psi_\nu^j(0)|0\rangle
=\int\frac{d^dp}{(2\pi)^d}\,\mbox{e}^{-ipx}\,S^L_{\mu\nu}(p)
\,\xi^{3/2,ij}\, 
\en
where
\eq\label{eq:Sp}
S_{\mu\nu}^L(p)=-\frac{1}{\md0-\not\! p\,+\Sigma^L(p)}\,(P^{3/2})_{\mu\nu}
+\cdots\, ,\quad\quad
\Sigma^L(p)=\not\! p\,\Sigma_1^L(p^2)+\md0\Sigma_2^L(p^2)\, ,
\en
and the ellipses stand for the terms with spin-$\frac{1}{2}$ projectors.
The poles of the propagator are given by the zeros of the denominator
\eq\label{eq:key0}
\md0-\not\! p\,+\not\! p\,\Sigma_1^L(p^2)+\md0\Sigma_2^L(p^2)=0\, .
\en
We further introduce the quantities
\eq
\tilde\Sigma_i^L(p^2)=\Sigma_i^L(p^2)-\mbox{Re}\,\Sigma_i(p^2)\, ,
\quad i=1,2\, ,
\en
where $\Sigma(p)=\not\! p\,\Sigma_1(p^2)+\md0\Sigma_2(p^2)$ denotes the 
self-energy part of the $\Delta$-resonance, calculated in the infinite volume
(note that this quantity is complex below the decay threshold).
Introducing now the physical $\Delta$-mass (the real part of the pole mass)
through the conventional
definition (in the infinite volume) 
\eq
m_\Delta=\md0(1+\mbox{Re}\,\Sigma_1(m_\Delta^2)
+\mbox{Re}\,\Sigma_2(m_\Delta^2))+O(p^4)\, ,
\en
it can be seen that, at the order we are working, Eq.~(\ref{eq:key0}) can
be rewritten by using the physical mass instead of the mass in the chiral limit
\eq\label{eq:key}
m_\Delta-\not\! p\,+\not\! p\,\tilde \Sigma_1^L(p^2)+m_\Delta\tilde\Sigma_2^L(p^2)=0\, .
\en
Performing the analytic continuation to Euclidian space
$p_\mu=(p_0,{\bf p})=(i\hat p_4,{\bf \hat p})$ and choosing 
the center-of-mass frame $\hat p_\mu=(\omega,{\bf 0})$, from Eq.~(\ref{eq:key})
we finally get
\eq\label{eq:key_f}
m_\Delta^2(1+\tilde\Sigma_2^L(-\omega^2))^2
+\omega^2(1-\tilde\Sigma_1^L(-\omega^2))^2=0\, .
\en

\subsection{\boldmath Calculation of the self-energy at $O(\epsilon^3)$}

Only pion-nucleon and pion-$\Delta$ loops, which are shown in Fig.~\ref{figure1},
 contribute to the quantities $\tilde\Sigma_i^L(-\omega^2)$. The counterterm
contribution is independent on $L$ and thus drops out from the final 
expression. One obtains
\eq\label{eq:S1}
\tilde\Sigma_i^L(-\omega^2)
=\tilde\Sigma_{i,N}^L(-\omega^2)+\tilde\Sigma_{i,\Delta}^L(-\omega^2)\, ,
\quad\quad i=1,2\, ,
\en
where
\eq\label{eq:Sn}
\tilde\Sigma_{1,N}^L(-\omega^2)=\frac{g_{\pi N\Delta}^2}{F^2}\,\biggl\{
\tilde W_2^N(-\omega^2)-\tilde W_3^N(-\omega^2)\biggr\}\, ,\quad\quad
\tilde\Sigma_{2,N}^L(-\omega^2)=\frac{g_{\pi N\Delta}^2}{F^2}\,\frac{m_N}{m_\Delta}\,
\tilde W_2^N(-\omega^2)
\en
and
\eq\label{eq:Sd}
\tilde\Sigma_{1,\Delta}^L(-\omega^2)&=&\frac{5g_1^2}{12F^2}\,\biggl\{
-\tilde T^\pi_0+(m_\Delta^2+\omega^2)\tilde W^\Delta_0(-\omega^2)
+(m_\Delta^2-\omega^2)\tilde W^\Delta_1(-\omega^2)
\nonumber\\[2mm]
&-&\frac{2}{3m_\Delta^2}\,(3m_\Delta^2+\omega^2)\tilde W^\Delta_2(-\omega^2)
-\frac{2}{3m_\Delta^2}\,(m_\Delta^2-\omega^2)\tilde W^\Delta_3(-\omega^2)
\biggr\}\, ,
\nonumber\\[2mm]
\tilde\Sigma_{2,\Delta}^L(-\omega^2)&=&\frac{5g_1^2}{12F^2}\,\biggl\{
-\tilde T^\pi_0+(m_\Delta^2+\omega^2)\tilde W^\Delta_0(-\omega^2)
-2\omega^2\tilde W^\Delta_1(-\omega^2)
\nonumber\\[2mm]
&-&\frac{4}{3}\,\tilde W^\Delta_2(-\omega^2)
+\frac{4\omega^2}{3m_\Delta^2}\,\tilde W^\Delta_3(-\omega^2)\biggr\}\, .
\en
In the above expressions, the quantity $\tilde T_0^\pi$ corresponds
to the pion tadpole and $\tilde W_i^N(-\omega^2)$,
$\tilde W_i^\Delta(-\omega^2)$ to the $\pi N$ and $\pi\Delta$ scalar loop
functions, respectively (note that the tadpoles emerge from the one-loop 
diagrams shown in Fig.~\ref{figure1} after simplifying the numerators).
An exact definition of these quantities is given below.
Further, in these 
expressions one may take $d=4$ because all ultraviolet divergences
are contained in the infinite-volume integrals which, at the end, are included
in the definition of the infinite-volume mass $m_\Delta$. Further, at the 
order we are working, the quantity $F$ can be 
replaced by the pion decay constant $F_\pi$.
 At the end, the 
self-energy part in a finite volume at $O(\epsilon^3)$ is expressed in 
terms of  physical quantities only.

In the calculations at  finite volume the loop integrals are replaced
by infinite sums over discrete lattice momenta 
${\bf k}_n=2\pi{\bf n}/L,~{\bf n}\in Z^3$ in spatial dimensions
(see, e.g.~\cite{Gasser:1987zq}).
To ease the notation, below we give the loop functions
with the infinite-volume part included. In order to arrive at the quantities
that enter Eqs.~(\ref{eq:Sn},\ref{eq:Sd}), one has first to isolate
the infinite-volume part and then subtract it, e.g.
\beq
\tilde T_0^\pi=T_0^\pi-T_0^\pi\bigr|_{L\to\infty}~.
\eeq
Consequently, the expressions that are given below, are ultraviolet divergent
and imply the application of some kind of  regularization.
Here we do not refer to a particular regularization procedure
explicitly, since
the divergent infinite-volume part will be always subtracted before
the actual calculations are performed.

The one-loop integrals that contribute to the self-energy of the
 $\Delta$-resonance, are listed below ($X=N,\Delta$). 
These are tadpole diagrams
\eq\label{eq:tadpoles}
&&T_0^\pi=\int_{-\infty}^\infty\frac{dk_4}{2\pi}\frac{1}{L^3}\sum_{{\bf n}}
\frac{1}{M_\pi^2+k_n^2}\, ,
\nonumber\\[2mm]
&&T_0^X=\int_{-\infty}^\infty\frac{dk_4}{2\pi}\frac{1}{L^3}\sum_{{\bf n}}
\frac{1}{m_X^2+k_n^2}\, ,
\nonumber\\[2mm]
&&\biggl(\delta_{\alpha\beta}
-\frac{\hat p_\alpha \hat p_\beta}{\hat p^2}\biggr)N_2^\pi
+\frac{\hat p_\alpha \hat p_\beta}{\hat p^2}\,T_2^\pi=
- \int_{-\infty}^\infty\frac{dk_4}{2\pi}\frac{1}{L^3}\sum_{{\bf n}}
\frac{k_\alpha k_\beta}{M_\pi^2+k_n^2}\, ,
\nonumber\\[2mm]
&&\biggl(\delta_{\alpha\beta}-\frac{\hat p_\alpha \hat p_\beta}{\hat p^2}\biggr)N_2^X
+\frac{\hat p_\alpha \hat p_\beta}{\hat p^2}\,T_2^X=
-\int_{-\infty}^\infty\frac{dk_4}{2\pi}\frac{1}{L^3}\sum_{{\bf n}}
\frac{k_\alpha k_\beta}{m_X^2+k_n^2}\, ,
\en
as well as the meson-baryon loop functions
\eq\label{eq:oneloop}
&&W_0^X(-\omega^2)=
\int_{-\infty}^\infty\frac{dk_4}{2\pi}\frac{1}{L^3}\sum_{{\bf n}}
\frac{1}{(M_\pi^2+k_n^2)(m_X^2+(\hat p-k_n)^2)}\, ,
\nonumber\\[2mm]
&&\hat p_\alpha W_1^X(-\omega^2)=
\int_{-\infty}^\infty\frac{dk_4}{2\pi}\frac{1}{L^3}\sum_{{\bf n}}
\frac{k_\alpha}
{(M_\pi^2+k_n^2)(m_X^2+(\hat p-k_n)^2)}\, ,
\nonumber\\[2mm]
&&\biggl(\delta_{\alpha\beta}
-\frac{\hat p_\alpha \hat p_\beta}{\hat p^2}\biggr)W_2^X(-\omega^2)
+\frac{\hat p_\alpha \hat p_\beta}{\hat p^2}\,S_2^X(-\omega^2)
= - \int_{-\infty}^\infty\frac{dk_4}{2\pi}\frac{1}{L^3}\sum_{{\bf n}}
\frac{k_\alpha k_\beta}
{(M_\pi^2+k_n^2)(m_X^2+(\hat p-k_n)^2)}\, ,
\nonumber\\[2mm]
&&(\delta_{\alpha\beta}\hat p_\sigma+\delta_{\sigma\alpha}\hat p_\beta
+\delta_{\beta\sigma}\hat p_\alpha)W_3^X(-\omega^2)
+\frac{\hat p_\alpha \hat p_\beta \hat p_\sigma}{\hat p^2}\,S_3^X(-\omega^2)
\nonumber\\[2mm]
&&\qquad\qquad\qquad\qquad\qquad
=\,\, - \int_{-\infty}^\infty\frac{dk_4}{2\pi}\frac{1}{L^3}\sum_{{\bf n}}
\frac{k_\alpha k_\beta k_\sigma}
{(M_\pi^2+k_n^2)(m_X^2+(\hat p-k_n)^2)}\, ,
\en
with the Euclidean 4-momentum $(k_n)_\mu=(k_4,{\bf k}_n)$.
After subtracting the infinite-volume piece, the following recurrence relations between
various functions can be obtained:
\eq
\tilde N_2^\pi&=&\frac{1}{3}\,(M_\pi^2 \tilde T_0^\pi-\tilde T_2^\pi)\, ,\quad\quad
\tilde N_2^X=\frac{1}{3}\,(m_X^2 \tilde T_0^X-\tilde T_2^X)\, ,
\nonumber\\[2mm]
\tilde W_1^X(-\omega^2)&=&\frac{\omega^2+m_X^2-M_\pi^2}{2\omega^2}\,
\tilde W_0^X(-\omega^2)-\frac{\tilde T_0^\pi}{2\omega^2}
+\frac{\tilde T_0^X}{2\omega^2}\, ,
\nonumber\\[2mm]
\tilde W_2^X(-\omega^2)&=&\frac{1}{12\omega^2}\,\biggl\{
\lambda(-\omega^2,m_X^2,M_\pi^2)\tilde W_0^X(-\omega^2)
-(\omega^2+m_X^2-M_\pi^2)\tilde T_0^\pi
-(\omega^2-m_X^2+M_\pi^2)\tilde T_0^X\biggr\}\, ,
\nonumber\\[2mm]
\tilde W_3^X(-\omega^2)&=&
\frac{\omega^2+m_X^2-M_\pi^2}{2\omega^2}\,\tilde W_2^X(-\omega^2)
-\frac{1}{6\omega^2}\,\biggl\{M_\pi^2\tilde T_0^\pi-\tilde T_2^\pi
-m_X^2\tilde T_0^X+\tilde T_2^X\biggr\}\, ,
\en
where $\lambda(x,y,z)=x^2+y^2+z^2-2xy-2yz+2zx$ denotes the usual triangle 
function.

\subsection{Calculation of the scalar integrals}

The calculation of the tadpole graphs is straightforward and is carried out
by using standard techniques (see, e.g.~\cite{AliKhan:2003cu,Beane:2004tw}). 
We demonstrate the method for the case of the pion tadpole.
Using dimensional regularization to tame the ultraviolet divergence
in Eq.~(\ref{eq:tadpoles}), one gets
\eq
T_0^\pi&=&
\frac{1}{L^3}\int_{-\infty}^\infty\frac{dk_4}{2\pi}\int d^{d-1}{\bf k}\,
\frac{1}{M_\pi^2+k_4^2+{\bf k}^2}\,\sum _{\bf n}
\delta^{d-1}\biggl({\bf k}-\frac{2\pi{\bf n}}{L}\biggr)
\nonumber\\[2mm]
&=&\int\frac{d^dk}{(2\pi)^d}\,\frac{1}{M_\pi^2+k_4^2+{\bf k}^2}
\sum_{{\bf j}}\mbox{e}^{iL{\bf k}{\bf j}}\, ,
\en
where the Poisson formula
\eq
\sum_{n=-\infty}^{+\infty}\delta(x-n)
=\sum_{n=-\infty}^{+\infty}\mbox{e}^{2\pi i nx}
\en
has been used to arrive at the second equality. Further, in this sum
the term with ${\bf j}=0$ corresponds to the infinite-volume integral.
Separating this term, we finally obtain
\eq
T_0^\pi&=&\int\frac{d^dk}{(2\pi)^d}\,\frac{1}{M_\pi^2+k_4^2+{\bf k}^2}
+\sum_{j\neq 0}\frac{1}{4\pi^2Lj}\int_0^\infty dk_4 \,
\mbox{e}^{-Lj\sqrt{M_\pi^2+k_4^2}}
\nonumber\\[2mm]
&=&T_0^\pi\biggr|_{L\to\infty}+\frac{M_\pi^2}{4\pi^2}\,
\sum_{j\neq 0}\frac{K_1(M_\pi Lj)}{M_\pi Lj}\, ,
\en
where $j=|{\bf j}|=\sqrt{j_1^2+j_2^2+j_3^2}$
and $K_\nu(z)$ denotes the modified Bessel function.

The final result for the tadpoles is given by
\eq\label{eq:tadpoles_exp}
&&\tilde T_0^\pi=\frac{M_\pi^2}{4\pi^2}\sum_{{\bf j}\neq 0}
\frac{K_1(M_\pi  Lj)}{M_\pi  Lj}\, ,\quad
\tilde T_0^X=\frac{m_X^2}{4\pi^2}\sum_{{\bf j}\neq 0}
\frac{K_1(m_X  Lj)}{m_X  Lj}\, ,
\nonumber\\[2mm]
&&\tilde T_2^\pi=-\frac{M_\pi^4}{4\pi^2}\sum_{{\bf j}\neq 0}
\frac{K_2(M_\pi  Lj)}{(M_\pi  Lj)^2}\, ,\quad
\tilde T_2^X=-\frac{m_X^4}{4\pi^2}\sum_{{\bf j}\neq 0}
\frac{K_2(m_X  Lj)}{(m_X  Lj)^2}\, .
\en
In the calculation of the meson-baryon loop functions one has to distinguish
between two cases. In the $\pi\Delta$ loop, the variable $-\omega^2$ is below
threshold. Using the Feynman parameterization, one may combine two 
denominators and then use the same technique as for the calculation of the
tadpole contribution. As a result, one gets
\eq\label{eq:Dloop}
\tilde W_0^\Delta(-\omega^2)&=&\frac{1}{8\pi^2}\int_0^1dx\sum_{{\bf j}\neq 0} 
K_0\biggl(Lj\sqrt{g_\Delta(x,-\omega^2)}\biggr)\, ,
\nonumber\\[2mm]
g_\Delta(x,-\omega^2)&=&(1-x)M_\pi^2+xm_\Delta^2+x(1-x)\omega^2\, .
\en
Note that for $-\omega^2$ close to $m_\Delta^2$ the function
$g_\Delta(x,-\omega^2)$ never vanishes in the integration region.

In contrast to the above example, the $\pi N$ loop can not be calculated by
using the same method, because the variable $-\omega^2$ can now be above
the decay threshold $\Delta\to N\pi$. To calculate this quantity, 
the following trick has been used. First, in order to avoid the ultraviolet 
divergence in the infinite sum over momenta, we have subtracted the integral
at some scale $\omega^2=\mu^2$ below threshold (one subtraction is enough
for the convergence, but double subtraction enables one to achieve 
faster convergence). Now in the subtraction terms one is allowed to use the same 
technique as in the calculation of the $\pi\Delta$ loop, because $-\mu^2$
is below threshold. This strategy is illustrated below in detail.
The quantity $\tilde W_0^N(-\omega^2)$ which we are looking for is 
split into several terms
\eq\label{eq:N1}
\tilde W_0^N(-\omega^2)&=&W_0^N(-\omega^2)-\mbox{Re}\,W_0^N(-\omega^2)\biggr|_{L\to\infty}=H_1(-\omega^2)+H_2(-\omega^2)+H_3(-\omega^2)\, ,
\nonumber\\[2mm]
H_1(-\omega^2)&=&\biggl\{W_0^N(-\omega^2)-W_0^N(-\mu^2)
+(\omega^2-\mu^2)\frac{d}{d\omega^2}
W_0^N(-\omega^2)\biggr|_{\omega^2=\mu^2}\biggr\}\, ,
\nonumber\\[2mm]
H_2(-\omega^2)&=&\biggl\{\tilde W_0^N(-\mu^2)+(\omega^2-\mu^2)
\frac{d}{d\omega^2}\tilde W_0^N(-\omega^2)\biggr|_{\omega^2=\mu^2}\biggr\}\, ,
\nonumber\\[2mm]
H_3(-\omega^2)&=&-\biggl\{\mbox{Re}\, W_0^N(-\omega^2)-W_0^N(-\mu^2)
+(\omega^2-\mu^2)\frac{d}{d\omega^2}W_0^N(-\omega^2)\biggr|_{\omega^2=\mu^2}
\biggr\}\biggr|_{L\to\infty}\, .
\en
The first term is the twice-subtracted infinite momentum sum, where the 
integration over $k_4$ is explicitly performed
\eq\label{eq:N2}
H_1(-\omega^2)&=&(\omega^2-\mu^2)^2\frac{1}{L^3}\sum_{{\bf n}}
\frac{E_N+E_\pi}{2E_N E_\pi}\,\frac{1}{\omega^2+(E_N+E_\pi)^2}\,
\frac{1}{(\mu^2+(E_N+E_\pi)^2)^2}\, ,
\nonumber\\[2mm]
E_N&=&\sqrt{m_N^2+{\bf k}_n^2}\, ,\quad\quad
E_\pi\,=\,\sqrt{M_\pi^2+{\bf k}_n^2}\, .
\en
The second expression corresponds to the subtraction term 
\eq\label{eq:N3}
H_2(-\omega^2)&=&\frac{1}{8\pi^2}\,\int_0^1dx\sum_{{\bf j}\neq 0}
\biggl(K_0\biggl(Lj\sqrt{g_N(x,-\mu^2)}\biggr)
\nonumber\\[2mm]
&-&(\omega^2-\mu^2)\frac{x(1-x)Lj}{2\sqrt{g_N(x,-\mu^2)}}\,
K_1\biggl(Lj\sqrt{g_N(x,-\mu^2)}\biggr)\biggr)\, ,
\nonumber\\[2mm]
g_N(x,-\mu^2)&=&(1-x)M_\pi^2+xm_N^2+x(1-x)\mu^2~.
\en
Furthermore, the remainder is included into the third term, which contains only
quantities evaluated in the infinite volume
\eq\label{eq:N4}
H_3(-\omega^2)&=&-\frac{B_\omega}{32\pi^2\omega^2}\biggl\{
\ln\frac{-\omega^2+m_N^2-M_\pi^2+B_\omega}{-\omega^2+m_N^2-M_\pi^2-B_\omega}
+\ln\frac{-\omega^2-m_N^2+M_\pi^2+B_\omega}{-\omega^2-m_N^2+M_\pi^2-B_\omega}
\biggr\}
\nonumber\\[2mm]
&+&\frac{B_\mu}{16\pi^2\mu^2}\biggl\{\arctan\frac{-\mu^2+m_N^2-M_\pi^2}{B_\mu}
+\arctan\frac{-\mu^2-m_N^2+M_\pi^2}{B_\mu}\biggr\}
\nonumber\\[2mm]
&-&\frac{\omega^2-\mu^2}{16\pi^2\mu^2}\biggl\{1
+\frac{(\omega^2-\mu^2)(m_N^2-M_\pi^2)}{2\omega^2\mu^2}\ln\frac{m_N^2}{M_\pi^2}
-\frac{\mu^2(m_N^2+M_\pi^2)+(m_N^2-M_\pi^2)^2}{\mu^2B_\mu}
\nonumber\\[2mm]
&\times&\biggl(\arctan\frac{-\mu^2+m_N^2-M_\pi^2}{B_\mu}
+\arctan\frac{-\mu^2-m_N^2+M_\pi^2}{B_\mu}\biggr)\biggr\}\, ,
\nonumber\\[2mm]
B_\omega&=&\lambda^{1/2}(-\omega^2,m_N^2,M_\pi^2)\, ,\quad\quad
B_\mu=\lambda^{1/2}(-\mu^2,m_N^2,M_\pi^2)\, .
\en

\subsection{Remarks}

\begin{itemize}

\item[a)]
In order to study baryon resonances, the volume should be taken much
larger than in the case of stable particles since for  excited states,
the volume-dependent effects decrease only as powers of $L$ and not by
an exponential law as for stable particles. More precisely, the parameter $L$
should be large enough, so that one could neglect all contributions of the
type $\exp(-\mbox{const}\cdot M_\pi L)$ 
(see, e.g.~\cite{LuescherI,Gasser:1987zq})
as compared to the corrections that
decrease according to the power law.

\item[b)]
The Lagrangians, given by Eqs.~(\ref{eq:L},\ref{eq:LN},\ref{eq:LND}) and
Eq.~(\ref{eq:LD}) do not contain  
the so-called ``off-shell'' pieces that do not 
contribute to observable quantities like the $S$-matrix elements or 
transition currents. They could in principle 
contribute to the self-energy of the $\Delta$, which is not an ``on-shell''
quantity. However, eliminating these terms at
the diagrammatic level corresponds to canceling one of the propagators in
the loop. The pertinent diagram turns into a tadpole, which vanishes 
exponentially. Therefore, for large values of $L$, where the exponential
factors can be neglected, the off-shell couplings do not contribute to the
self-energy.

\item[c)]
The same line of reasoning can be used to show that the contribution
from the spin-$\frac{1}{2}$ components of the $\Delta$-propagator is irrelevant
at  finite volume.

\item[d)]
As seen from Eq.~(\ref{eq:Dloop}), the whole $L$-dependent part of the 
contribution from the $\pi\Delta$ loop is exponentially suppressed
at  large $L$.
The same is true for all tadpoles. Even if we have retained them in the final
expressions for completeness, the $L$-dependent part thereof can be safely 
neglected at any stage of the calculation. Indeed, we have checked numerically
that the contribution from the $\pi\Delta$ loop is very small and does not
affect the results.

\item[e)]
Covariant SSE calculations in the baryon sector at an infinite 
volume are performed by e.g. using infrared regularization (IR), which leads to a 
consistent power counting in the presence of the (large) baryon mass.
This method has been also applied in the calculations at finite 
volume (see, e.g.~\cite{AliKhan:2003cu}). The procedure, adopted in that work,
amounts to merely extending the integration interval over the 
pertinent Feynman parameter from $[0,1]$ to $[0,\infty[$, in a complete
analogy to what is done at an infinite volume. 
In other words, in the context of the present problem it is equivalent to
setting the $N$ and $\Delta$ tadpoles to $0$ in Eq.~(\ref{eq:tadpoles_exp})
and to extending the integration over $dx$ in Eq.~(\ref{eq:Dloop}) from $0$
to $\infty$. It is, however, not immediately clear how this procedure can be 
generalized to the
case when $-\omega^2$ is above threshold, see 
Eqs.~(\ref{eq:N1},\ref{eq:N2},\ref{eq:N3},\ref{eq:N4}).
Our method is different
from the one described above. Namely, we note first that applying IR 
in an infinite volume is equivalent to using ordinary dimensional
regularization and changing the renormalization prescription.
Next, going to a finite volume implies using the same
Lagrangian, while replacing integrals by sums in the loops. 
In other words, 
our prescription reduces to using IR only in the infinite-volume
self-energy $\Sigma(p)$, whereas in the finite-volume piece
$\tilde \Sigma^L(p)=\Sigma^L(p)-\Sigma(p)$  ordinary dimensional 
regularization has been used. 
It is clear that in our case, unlike Ref.~\cite{AliKhan:2003cu},
 the counterterms, corresponding to the
above-mentioned additional renormalization, are $L$-independent.

In the case of a stable particle 
changing the prescription from one to 
another amounts to introducing corrections which vanish as
$\exp(-\mbox{const}\cdot m_XL)$ with $X=N,\Delta$ at large $L$. 
This happens, in particular,
for the volume-dependent nucleon mass, considered in 
Ref.~\cite{AliKhan:2003cu}. It is clear that in this case the difference
between the two prescriptions is physically irrelevant. It remains to be seen,
what kind of statement can be made in case of an unstable particle.

\end{itemize}

\section{Numerical results}
\label{sec:numerics}

Solving Eq.~(\ref{eq:key_f}) for different values of $L$ numerically
we get the dependence of the energy levels  $E_n(L)$ of the system, placed
in a Euclidean box of  size $L$. At this order, the energy levels
depend on two parameters, namely the mass $m_\Delta$ and
the coupling constant  $g_{\pi N\Delta}$. 
The term containing
the coupling constant $g_1$ is exponentially suppressed and we assume that
$m_N,M_\pi,F_\pi$ are determined independently in the same lattice 
calculations. We wish
to investigate whether one can determine $m_\Delta$ and $g_{\pi N\Delta}$
at a reasonable accuracy from a fit to the energy levels.

\begin{figure}[t]
\begin{center}
\includegraphics[width=10.cm]{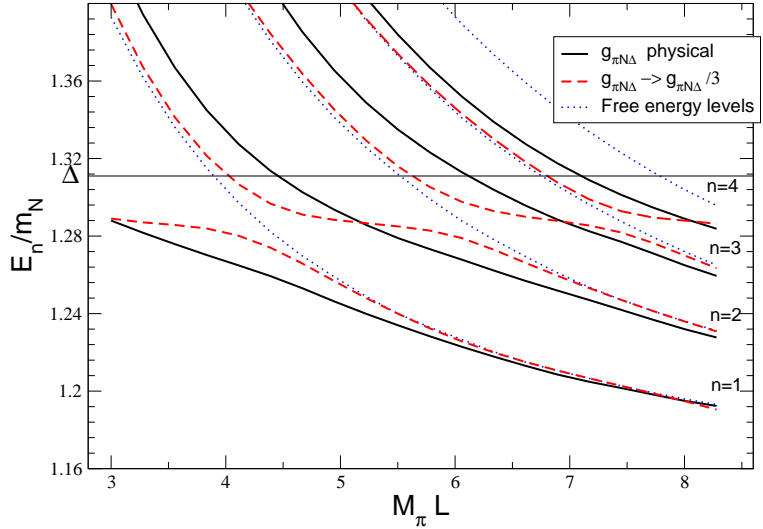}
\end{center}
\caption{The dependence of the energy levels in a finite box on the box 
size $L$ for different values of the coupling constant $g_{\pi N\Delta}$.
The avoided level crossing, which is clearly seen at small values of
$g_{\pi N\Delta}$ (dashed lines), is washed out for the physical value 
of this coupling 
constant (solid lines). For comparison, we also display the free energy levels
(dotted lines). It is seen that the energy levels in the presence of 
the interaction interpolate between different free energy levels. 
As expected, an abrupt change emerges in the vicinity of the resonance energy
(the resonance position corresponds to the solid horizontal line).}
\label{figure2}
\end{figure}

The results of calculations of the energy levels for the physical values of
all parameters are shown in Fig.~\ref{figure2} (solid lines). We use the
following values for the particle masses: $M_\pi=140~\mbox{MeV}$, $m_N=940~\mbox{MeV}$
and $m_\Delta=1210~\mbox{MeV}$ (the real part of the pole mass)
and for the pion decay constant  $F_\pi=92.4~\mbox{MeV}$. 
Further, we fix the physical value of the $\pi N\Delta$ coupling
constant as $g_{\pi N\Delta}=1.2$ (at $M_\pi=140~\mbox{MeV}$).
This corresponds to the value $g_\Delta^2/4\pi=13.5~\mbox{GeV}^{-2}$ 
($g_{\pi N\Delta}=F_\pi g_\Delta$) obtained by using the pole 
approximation in dispersion relations, see Ref.~\cite{Hoehler}. 
Finally, we take $g_1=2.0$ from Ref.~\cite{Bernard:2005fy} 
(nothing changes if we take $g_1=0$).
As we see, the coupling of the $\Delta$ to the $\pi N$-system 
is so strong that the
nice structure with the avoided level crossing has been almost completely
washed out. It will resurface again, if the input value of $g_{\pi N\Delta}$
is drastically 
reduced by hand (dashed lines in the same figure). This property, however,
can not be meaningfully used in the fitting procedure. It is clear that, 
in order to perform the fit, another 
strategy, not linked to the identification of the resonance energy
from the position of the avoided level crossing,  should be looked for.

\begin{figure}[t]
\begin{center}
\includegraphics[width=10.cm]{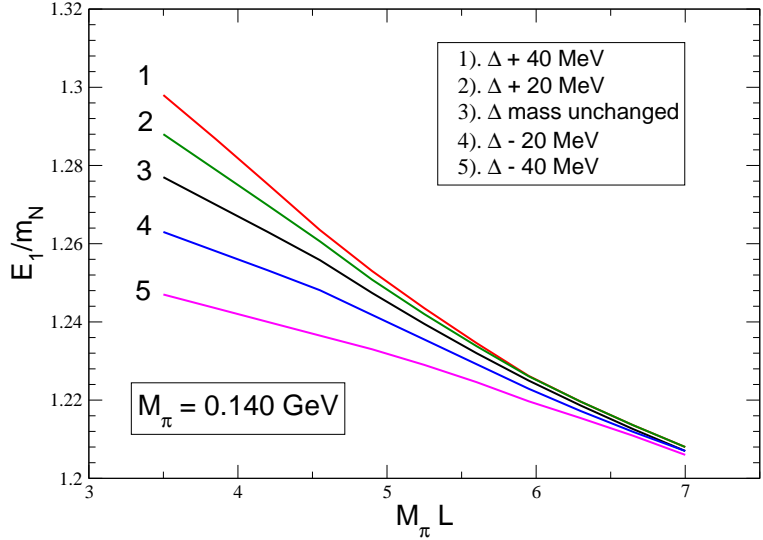}
\end{center}
\caption{The lowest energy level for different input values of $m_\Delta$ 
and the physical values of $g_{\pi N\Delta}$ and $M_\pi$.}
\label{Deltamass}
\end{figure}

In order to find such a strategy, we continue to study the structure
of the energy levels, as well as the dependence 
on  all available parameters. In particular, 
we start with varying $m_\Delta$
keeping all other parameters fixed. Consider for instance the lowest energy
level shown in Fig.~\ref{Deltamass},  where the dependence of the energy 
eigenvalue on $L$ is plotted for different input value of $m_\Delta$.
It is seen that the curves are almost linear in the variable
$\hat L=M_\pi L$ from the interval shown, with the tangent that monotonically
decreases with decreasing $m_\Delta$. This property can be used
for extracting $m_\Delta$.
More precisely, one has  to measure $E_1$ on the lattice for  different $\hat L$
and fit the curve, treating $m_\Delta$ as a free parameter.
The lattice data decide which of the curves
in Fig.~\ref{figure3} has to be chosen. Since each curve corresponds to a 
particular value of the $m_\Delta$, the fit to the lowest level determines
this parameter unambiguously (provided $g_{\pi N\Delta}$ is known).

\begin{figure}[t]
\begin{center}
\includegraphics[width=10.cm]{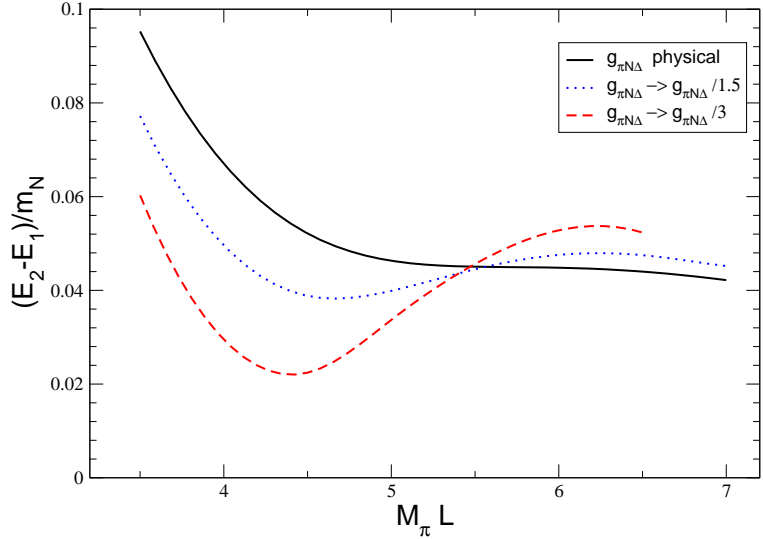}
\end{center}
\caption{The difference of the first two energy levels for different values of 
$g_{\pi N\Delta}$ and the physical value of $M_\pi$. For small values of the 
coupling constant there is a dip,
corresponding to the avoided level crossing. A plateau is clearly visible even
for the physical value of $g_{\pi N\Delta}$.} 
\label{figure4}
\end{figure}

Next, we consider the possibility of the determination of the $\pi N\Delta$
coupling constant. To this end, one has to find a quantity that could be 
maximally sensitive to $g_{\pi N\Delta}$ and try to extract this coupling 
by a fitting procedure. Bearing in mind the level structure in the presence
of a very narrow resonance (avoided level crossing), we expect that the 
difference of two energy levels $E_2(L)-E_1(L)$ can strongly depend
on the width of the resonance. In Fig.~\ref{figure4} we plot the quantity
$E_2(L)-E_1(L)$ against the  variable $\hat L$
at a fixed value of $m_\Delta$ and varying the parameter
$g_{\pi N\Delta}$. The avoided level crossing in this quantity -- 
at small values of the parameter $g_{\pi N\Delta}$ -- is seen as
a sharp minimum near the value of $\hat L$ where the crossing 
takes place and the
value of the function at the minimum determines the width.
As evident from Fig.~\ref{figure4}, even at the physical value of 
$g_{\pi N\Delta}$
one may observe a remnant of the avoided level crossing -- a plateau, which
disappears if $g_{\pi N\Delta}$ increases further. Given a rather pronounced
dependence of the quantities plotted in Fig.~\ref{figure4} on the input value of
$g_{\pi N\Delta}$, one may expect that fitting would allow one to
determine this coupling constant with a reasonable accuracy.

We are now in a position to describe our proposal for determining 
the parameters of the $\Delta$-resonance $m_\Delta$ and $g_{\pi N\Delta}$ 
from the lattice data, which is expected to work, even if the width of the
resonance is not very small. In brief, we propose to fit the first few
energy levels, measured on the lattice, to the  
calculated energy levels, which are parameterized by the free parameters
$m_\Delta$ and $g_{\pi N\Delta}$. Further, it could be advantageous
to carry out this procedure iteratively. Fix first the decay
constant to some input value and determine the
$\Delta$-mass from the $\hat L$-dependence of
the lowest energy level, see Fig.~\ref{Deltamass}. 
With the newly determined $\Delta$-mass plot the difference
$E_2(L)-E_1(L)$ and fit the parameter $g_{\pi N\Delta}$ to the data,
see Fig.~\ref{figure4}. Repeat the procedure until convergence
is achieved.

Finally, we wish to comment on the dependence of the energy levels on the
other parameter which is at our disposal, namely 
the quark (pion) mass. Since this structure depends only on
the physical nucleon and $\Delta$ masses, in the present (exploratory)
study of the problem we have restricted ourselves to the $O(\epsilon^2)$
expressions of the baryon masses
\eq
m_N=m_N^{\rm phys}-4c_1(M_\pi^2-(M_\pi^{\rm phys})^2)\, ,\quad\quad
m_\Delta=m_\Delta^{\rm phys}-4a_1(M_\pi^2-(M_\pi^{\rm phys})^2)
\en
and take $c_1=-0.9~\mbox{GeV}^{-1}$, 
$a_1=-0.3~\mbox{GeV}^{-1}$~\cite{Bernard:2005fy} (note that $c_1$ and $a_1$
enter only through the nucleon and $\Delta$ masses in the infinite volume, 
so the choice of particular numerical values for these constants does not
affect our conclusions).
Moreover, we neglect the pion mass dependence of the coupling constant
$F_\pi$, since the pertinent correction arises at higher
order. The above simple case perfectly models the real situation: the mass
difference between $m_\Delta$ and $m_N+M_\pi$ monotonically decreases
with the increase of $M_\pi$ and vanishes at around $M_\pi=210~\mbox{MeV}$.
Below threshold, the structure of the energy levels and the dependence
on the parameters $m_\Delta$, $g_{\pi N\Delta}$ is similar to the case
with $M_\pi=140~\mbox{MeV}$.
After crossing the threshold from below, 
the $\Delta$ is stable and one expects that the 
finite-volume corrections to the lowest energy level get exponentially
suppressed. 

Fig.~\ref{figure3} clearly illustrates this pattern. In this 
figure, the lowest-order energy level is plotted at three different values
of the pion mass and the physical value of the decay constant. As we see,
the curve is rather smooth in all cases and monotonically flattens as
the decay threshold is approached from below. This property can be
discussed in a more quantitative fashion.
Namely, fitting the level energy in Fig.~\ref{figure3}
$\hat E_1=E_1/m_N$ to the variable $\hat L$ within the
interval $3\leq \hat L\leq 8.3$ with a linear
function $\hat E_1=A+B\hat L$, for the different values of the mass gap
$\omega_0=m_\Delta-m_N-M_\pi$ we get:
$(A,B)=(1.34,-0.019)$ for $\omega_0=130~\mbox{MeV}$,
$(A,B)=(1.30,-0.009)$ for $\omega_0=78~\mbox{MeV}$,
$(A,B)=(1.25,-0.006)$ for $\omega_0=20~\mbox{MeV}$.
It is seen that, while the parameter $A$ remains almost stable, the linear
coefficient decreases monotonically. It is important to observe that 
there is no sign of irregularity in the quantity $E_1(L)$, when one crosses the threshold.
Finally, we would like to mention that here the distance at which the
chiral extrapolation takes place is not bound from below and
could be taken smaller than the distance to the decay threshold.
Consequently, the extrapolation error could be reduced.

\begin{figure}[t]
\begin{center}
\includegraphics[width=10.cm]{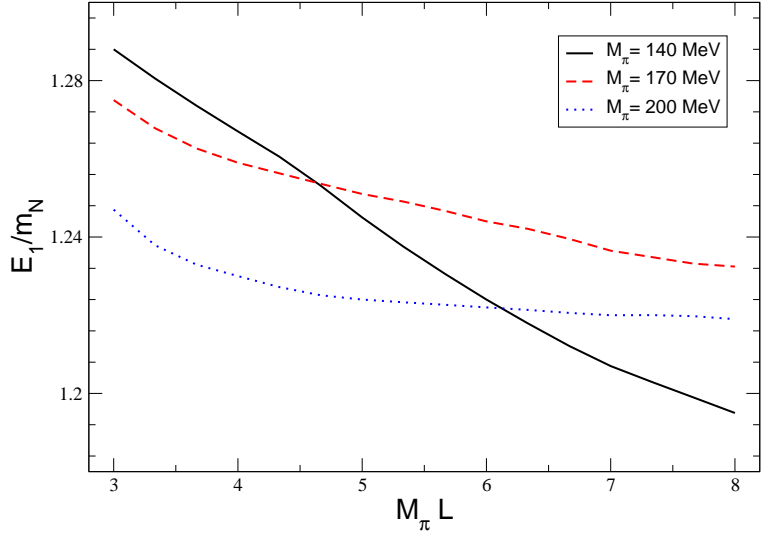}
\end{center}
\caption{The lowest energy level for different values of $M_\pi$ 
and the physical value of $g_{\pi N\Delta}$. The curves correspond to
the following values of $\omega_0\doteq m_\Delta-m_N-M_\pi$:
a) solid line: $\omega_0=130~\mbox{MeV}$,
b) dashed line: $\omega_0=78~\mbox{MeV}$,
c) dotted line: $\omega_0=20~\mbox{MeV}$.
}
\label{figure3}
\end{figure}

\section{Analytic parameterization}
\label{sec:analytic}

The dependence of the energy levels $E_n(L)$ on the parameters $m_\Delta$, 
$g_{\pi N\Delta}$ and $L$ is given by the numerical solution of Eq.~(\ref{eq:key_f}). 
If one implements the above fitting procedure in 
practice, it would be useful to have a simplified algebraic expression for 
the energy levels, where the dependence on the parameters is explicit. 
Below we demonstrate how such an expression can be derived.
Starting from Eqs.~(\ref{eq:S1},\ref{eq:Sn},\ref{eq:Sd}), we 
neglect all contributions that are exponentially suppressed in $L$. Namely, the
$\pi\Delta$ loop is neglected altogether as well as various tadpoles.
The functions $\tilde \Sigma_i(-\omega^2)$, which appear in the self-energy 
part, are then
proportional to the loop function $\tilde W_0^N(-\omega^2)$, defined
in Eq.~(\ref{eq:oneloop}). Then, Eq.~(\ref{eq:key_f})  simplifies to
\eq
m_\Delta-\sqrt{-\omega^2}=-\frac{g_{\pi N\Delta}^2}{F^2}\,
\frac{1}{2\sqrt{-\omega^2}}\,
\biggl\{\bigl(\sqrt{-\omega^2}+m_N\bigr)^2-M_\pi^2\biggr\}
\frac{\lambda(-\omega^2,m_N^2,M_\pi^2)}{12\omega^2}\,\tilde W_0^N(-\omega^2)\, .
\en
 Further, since the $\Delta$ is a $P$-wave state,
the lowest singularity in the self-energy
corresponds to the contribution of the term with ${\bf n}^2=1$ (it can be 
easily checked that the singularity at ${\bf n}^2=0$ cancels with
the factor $\lambda(-\omega^2,m_N^2,M_\pi^2)$ in the numerator).
Isolating the singularity at ${\bf n}^2=1$ in the function 
$\tilde W_0^N(-\omega^2)$, one may write
\eq
\tilde W_0^N(-\omega^2)=\frac{6}{L^3}\frac{E^{(1)}}{2E_N^{(1)}E_\pi^{(1)}}\,
\frac{1}{\omega^2+(E^{(1)})^2}+
\tilde R_0^N(-\omega^2)\, ,
\en
where
\eq
E_N^{(1)}=\sqrt{m_N^2+(2\pi/L)^2}\, ,\quad\quad
E_\pi^{(1)}=\sqrt{M_\pi^2+(2\pi/L)^2}\, ,\quad\quad
E^{(1)}=E_N^{(1)}+E_\pi^{(1)}~,
\en
and the function $\tilde R_0^N(-\omega^2)$ is regular in the vicinity
of $\omega^2=-(E^{(1)})^2$. Further, Eq.~(\ref{eq:key_f}) can be rewritten as
\eq
&&m_\Delta-\sqrt{-\omega^2}=\frac{g(L)}{E^{(1)}-\sqrt{-\omega^2}}
+r(\sqrt{-\omega^2})\, ,
\en
where
\eq
g(L)=\frac{g_{\pi N\Delta}^2}{F^2}\,\frac{1}{16 (E^{(1)})^3}\,
\biggl\{(E^{(1)}+m_N)^2-M_\pi^2\biggr\}
\lambda((E^{(1)})^2,m_N^2,M_\pi^2)\,\frac{1}{L^3 E_N^{(1)}E_\pi^{(1)}}
\en
and the function $r(\sqrt{-\omega^2})$ is regular at $\sqrt{-\omega^2}=E^{(1)}$.

\begin{figure}[t]
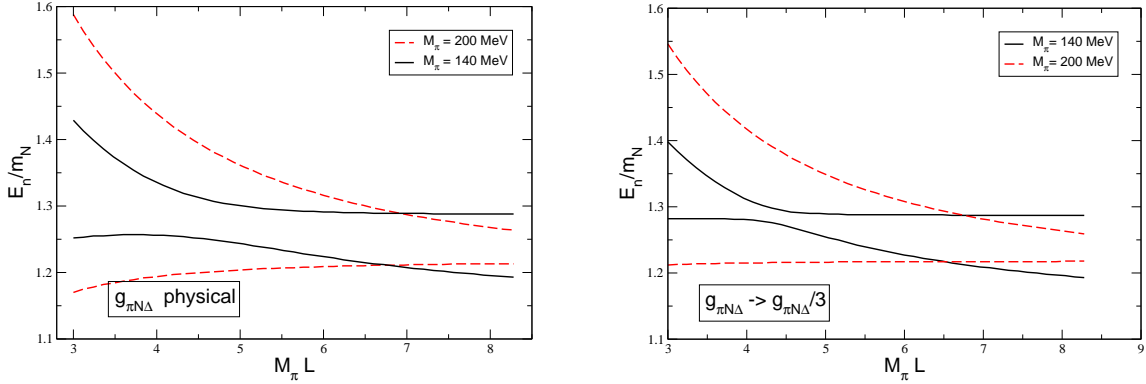

\begin{center}
\includegraphics[width=7.cm]{sqrt1-1.2.eps}\hspace*{1.cm}
\includegraphics[width=7.cm]{sqrt2-1.2.eps}
\end{center}
\caption{Approximate solution for the first two energy levels, Eq.~(\ref{eq:sqrt}),
for physical (left panel) and reduced (right panel) values 
of the coupling constant $g_{\pi N\Delta}$. The approximate solution 
qualitatively reproduces the features of the exact solution. Namely, the
avoided level crossing is clearly visible in the right panel, as well as an
almost stable $\Delta$-state at $M_\pi=200~\mbox{MeV}$ (left and right panels).}
\label{figure5}
\end{figure}

If now we use the approximation $r(\sqrt{-\omega^2})=0$, we arrive at a 
quadratic equation, whose solution 
\eq\label{eq:sqrt}
E_{1,2}(L)=
\frac{1}{2}\,\biggl\{m_\Delta+E^{(1)}\pm
\sqrt{(m_\Delta-E^{(1)})^2+4g(L)}\biggr\}
\en
gives the position of the first two energy levels.
In Fig.~\ref{figure5} these levels are plotted at the physical value of 
$g_{\pi N\Delta}$, as well as for the case of a smaller value. It is amusing
that such a simple solution reproduces the gross features of the exact solution
quite well. For example, the avoided level crossing is clearly visible
at smaller values of $g_{\pi N\Delta}$. Finally, we note that in
Ref.~\cite{Wiese} a very similar 
equation is derived in a simple two-channel quantum-mechanical model.

The approximate solution, given by Eq.~(\ref{eq:sqrt}), 
contains an explicit dependence on the parameters $L$, $m_\Delta$ and 
$g_{\pi N\Delta}$.
For this reason, it is  easier to use it for performing the fit 
to the lowest energy levels.
Of course, from the quantitative point of view,
the quality of the approximation is still not satisfactory.
It is however obvious that the accuracy
can be systematically improved by including the next nearby
singularities as well as regular contributions. We do not display the pertinent
formulae here. It is clear
that, at the end, the accuracy of the analytic parameterization, 
which is used to analyze the real lattice data, should be matched 
to the precision of this data.

\section{Conclusions}
\label{sec:concl}

\begin{itemize}

\item[a)]
In this paper we present the results of calculations of the pole structure
of the correlator of two $\Delta$-fields in a finite Euclidean box. 
It has been argued that, calculating the first few energy levels 
$E_n(L)$ in terms of the resonance parameters $m_\Delta$ and $g_{\pi N\Delta}$
within the SSE and fitting the lattice data at finite
volume, an extraction of these parameters at a reasonable accuracy may be 
possible. This statement constitutes the main result of the present paper.

\item[b)]
The main question that remains is, how the result will be affected
by higher-order corrections in the chiral expansion. It can be for instance 
shown that, parameterizing
pion-nucleon scattering matrix by the pure $s$-channel $\Delta$-pole term,
Eq.~(\ref{eq:key_f}) can be rewritten in a form similar to the L\"uscher's
master formula, which expresses the displacement of an energy level through the
scattering phase shift. The higher-order corrections in the SSE contribute
to the non-resonant background and are expected to be moderate. Of course,
such heuristic arguments can not be a real substitute of explicit 
calculations. Note that such calculations at $O(\epsilon^4)$ are already
in progress
and the results will be reported elsewhere, including a detailed error analysis
in SSE at this order~\cite{Hoja}.
One expects that these calculations shed light on the question of the convergence
of the chiral expansion.

\item[c)]
Our approach is closely related to the method proposed originally
by L\"uscher and developed in number of subsequent 
publications~\cite{LuescherII,Luescher_torus,Luescher_rho,Houches,Wiese,Beane:2003yx}. 
Note, however, that the L\"uscher formula, relating the energy levels of
a system in a finite volume to the scattering phase shifts, is valid in 
general beyond the chiral expansion, thus avoiding the above-mentioned
problem of  convergence. 
On the other hand, our result contains an explicit parameterization
of the energy levels in terms of $m_\Delta$, $g_{\pi N\Delta}$ and can be used,
in addition, to study the quark mass dependence of the energy levels.
This is important because the first lattice data which will appear 
below $\pi N$ threshold, will probably still correspond to the pion mass 
higher than the physical value.

We now note that, 
due to the condition $M_\pi L\gg 1$, the characteristic 3-momenta
of the system $p\ll M_\pi$ and 
the non-relativistic approach must be applicable
-- the processes with a mass gap $\sim M_\pi$ or higher are suppressed
exponentially. One therefore expects that, for a sufficiently large $L$
these two approaches overlap and the results are complementary
to each other. At present, we are investigating the problem in detail
within non-relativistic EFT, aiming
to explicitly demonstrate this relationship
that, in turn, will enable one to choose an optimum strategy for determining
the resonance parameters from the lattice data in the future~\cite{Lage}.

\end{itemize}

\bigskip

\noindent {\it {\bf Acknowledgments}:}\\[0.3em]
The authors would like to thank G.~Colangelo, J.~Gasser, C.~Haefeli,
M.~Savage, G.~Schierholz, R.~Sommer and U.-J.~Wiese
 for interesting discussions.

\baselineskip 12pt plus 1pt minus 1pt

\end{document}